\documentclass{PoS}

\PoS{PoS(BDMH2004)}
\usepackage{psfig}

\ShortTitle{Numerical heating in galaxy formation simulations}

\title{Cosmological particle-based simulations of galaxy formation; 
numerical loss of angular momentum and disk heating}

\author{\speaker{Lucio Mayer}\\
        Inst. for Theoretical Physics, Univ. of Zürich, Switzerland\\
        E-mail: \email{lucio@physik.unizh.ch}}

\abstract{

We discuss the role that numerical effects have in 
cosmological N-Body/SPH simulations of disk galaxy formation. We show
that the resolution of current state-of-the-art calculations,
about  $10^5$ SPH and dark matter  particles
within the virial radius of a Milky Way-sized halo at $z=0$,
is just enough to avoid artificial losses of angular momentum 
and dramatic numerical disk heating during the phase of disk assembly. 
Instead both effects will still be an issue 
at earlier epochs, when the progenitors of the final 
galaxies are resolved by significantly less than $10^5$ particles. 
 This must be at least partially responsible for why simulations
show a ubiquitous low angular momentum spheroid that is too massive compared 
to that of typical disk galaxies.}

\FullConference{Baryons in Dark Matter Halos\\
                 5-9 October 2004\\
                 Novigrad, Croatia}

\begin{document}

\section{Introduction}

A remarkable problem of current galaxy formation simulations in a LCDM
Universe is their inability to produce a disk dominated galaxy.
The latest numerical simulations 
can form disks with reasonable sizes ([1],[2],[3], [4]), but no simulation 
has managed to form a bulgeless galaxy with a realistic, kinematically cold 
thin stellar disk. 
A large fraction of the stellar mass of the simulated galaxies always comes in 
the form of a low angular momentum, kinematically hot component resembling 
observed stellar bulges. Traditionally loss of angular momentum of progenitor
lumps through dynamical friction has been usually blamed as the cause
of excessive angular momentum loss.
Feedback from supernovae has been invoked to maintain 
the gas in a hot, diffuse phase and reduce the effect of dynamical friction. 
However the physics of feedback is poorly understood and thus its modeling 
is highly uncertain. But to what extent can we trust the results of the 
current simulations aside from the feedback issue? 
It is somehow worrisome that so far essentially only one numerical 
technique, SPH, has been used to model the hydrodynamics in cosmological 
simulations of disk galaxy formation.  While comparisons with different 
approaches, for example with adaptive mesh codes, are mandatory, an
obvious step to take is trying to understand the role of numerical effects 
in the current simulations ([5],[6]).
A galaxy formation simulation is extremely complex. Just repeating the
same simulation with different mass and/or force resolution to seek
convergence can be a daunting task if it is not combined with simpler
experiments aimed at testing individual aspects of one 
simulation.

\begin{centering}
\begin{figure}
\psfig{file=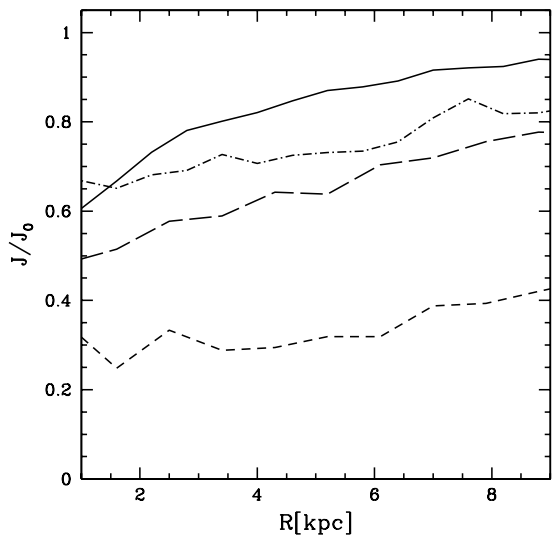,width=3.5truein,height=3.5truein}
\caption{Angular momentum loss in an isolated
disk galaxy model with structural parameters as our $\Lambda$CDM run
at z = 0.6. This model was run for 6 Gyrs ( equivalent to the present
time).  The y axis shows the fractional angular momentum loss for all
the baryonic material in the disk as a function of radius.  Continuous
line:N$_{DM}$ = 100000, Nstar=200000, Ngas=5000 (same as in the cosmo
run).  Dotted short dashed: N$_{DM}$
and Nstar reduced by a factor of five. Long dashed: Nstars reduced by
a factor of 25, the other components unchanged. Short dashed:N$_{DM}$= 4000,
 namely reduced by another factor of 5), the other components unchanged. At
the lowest resolution the disk undergoes the catastrophic angular
momentum loss reported in early simulations.}
\end{figure}
\end{centering}

\begin{centering}
\begin{figure}
\psfig{figure=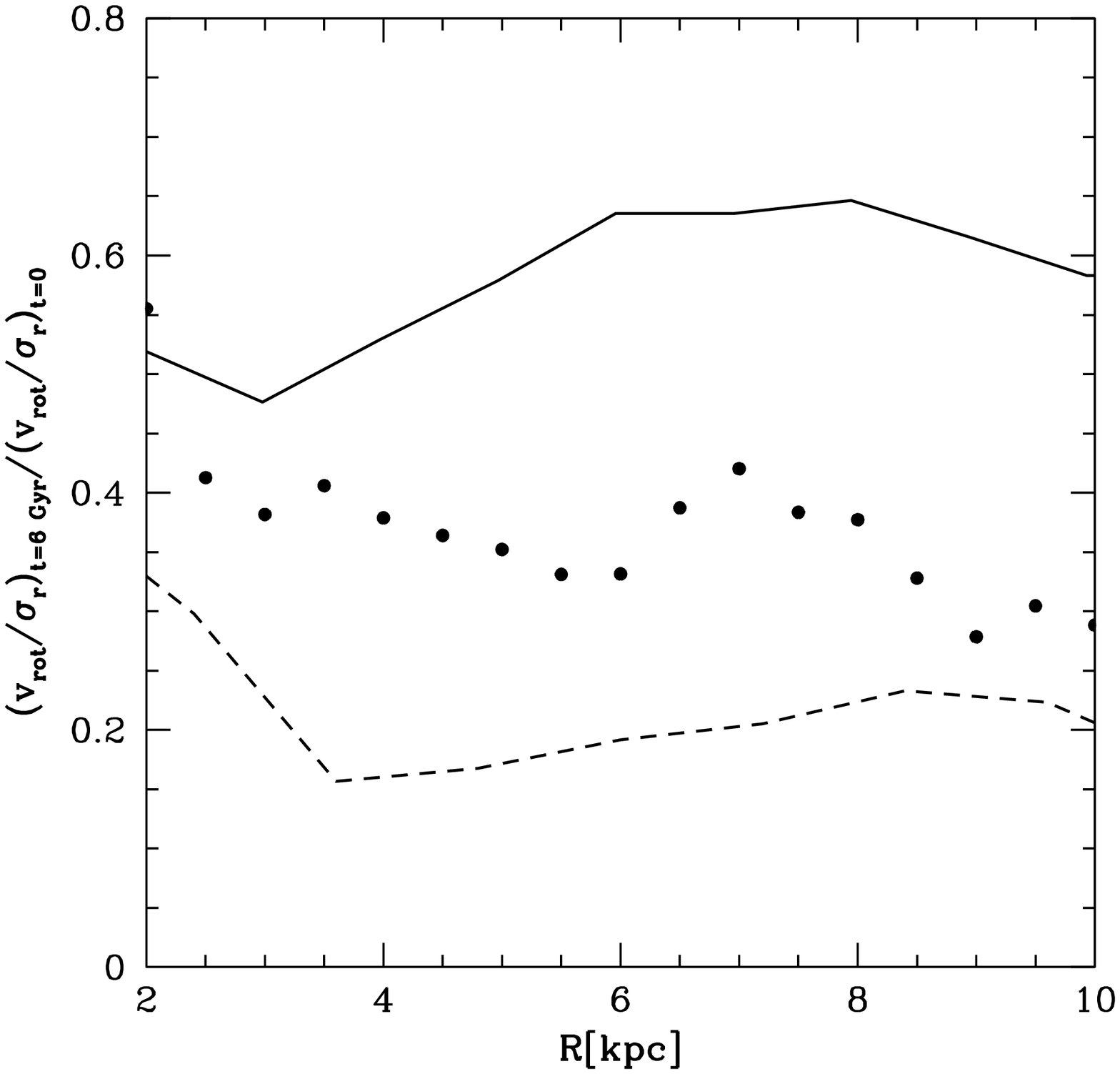,width=3.2truein,height=3.2truein}
\caption{Radial profile of $v_{rot}/\sigma$ after $\sim$ 6 Gyr of evolution
for the hi-res (solid line) and the low-res (dashed line) isolated galaxy 
test (see text), normalized to the profile of  $v_{rot}/\sigma$ at $t=0$.
The filled dots show the same profile for the galaxy in the LCDM simulation at
$z=0$ normalized to that at $z=0.6$.}
\end{figure}
\end{centering}

\section{Kinematically hot disks and numerical loss of angular momentum}

Disk dominated galaxies should form in halos that suffered the last major 
merger several Gyr before the present epoch. However, simulations
done during the past decade ([7]) typically used to study disk
formation without first selecting objects with quiet merging histories.
In addition to insufficient numerical resolution, this is certainly one 
reason why disks were found to be an order of magnitude too small compared 
to the observed ones. Recent simulations do take into account the
merging history. For example, [2] selected a galaxy-sized halo with
the last major merger occurring at $z=2.5$ in a large box (using the 
concordance LCDM model) and carried out a ``renormalized'' run using 
increased resolution in a  region of about 1 Mpc around such halo; at $z=0$
there are $\sim 1 \times 10^5$ dark matter particles and 
$\sim 6 \times 10^5$ between gas and star particles within the virial 
radius of the selected system (the force resolution was 1 kpc).
At $z=0$ the galaxy has a disk of nearly 16 kpc in size, 
where the disk edge is defined to be at the radius where
stars cease to be predominantly supported by rotaton.
Nevertheless, Figure 4 of [2] highlights two problems.
One is the presence of a massive spheroid, that produces a central peak 
in the rotation curve much more pronounced than that of the Milky Way or M31, 
the other one is the fact that $v_{rot}/\sigma$, namely the ratio
between rotation and velocity dispersion of the disk stars, is everywhere
a factor of  2-3 higher than that of a typical large disk galaxy 
($v_{rot}/\sigma > 5$). As a result the galaxy resembles 
the Sombrero galaxy rather than an Sb galaxy like the Milky Way.
While the spheroid is mostly 
made by old stars that were born in the progenitor lumps or during the
last major merger, the disk forms inside out mostly from the smooth accretion 
of halo gas [8] that cools and gradually settles into centrifugal support. 
Therefore the formation of the disk component is to a large extent not 
hierarchical. Certainly there are several satellites orbiting in the main
halo; these are partially disrupted by the tides of the primary, yet they 
never  merge with the disk because dynamical friction times are too long, 
and the amount of stars they lose to the disk is a negligible fraction
of its mass. 
Hence it is sensible to explore numerical effects during the 
disk assembly phase by setting up controlled experiments with no
cosmological initial conditions. One example of this approach can be found 
in [9] (these proceedings), which presents simulations
of disk formation from cooling of gas in an isolated spinning NFW halo.
Another example is described in [2], where multi-component
equilibrium disk models were used to estimate the angular momentum loss induced
by numerical effects. One of such effects is collisions of 
massive halo particles with much lighter gas and star particles. This is
just numerical two-body relaxation in a system with different particle
species [10].

Two-body heating can artificially randomize and increase the
kinetic energy of a (kinematically) cold rotating stellar disk 
[11]. It can also increase the thermal energy of the gas 
and affect the amount of cold gas, and thus stars, that ends up in the 
disk component. [12] showed that the second effect is under control 
when the number of halo particles is above $\sim 1000$, hence it is
negliglible in current simulations (except maybe in the earliest structures
that form at high redshift). To test the first effect
we built a multi-component galaxy model (with stellar disk, bulge and 
gaseous disk embedded in a NFW adiabatically contracted
dark halo) having structural parameters that match very closely
those of the galaxy in the LCDM simulation at $z=0.6$ (at this time most of 
the disk mass is already in place). We evolved the model for about 6 Gyr, 
roughly the time span between $z=0.6$ and $z=0$ in a LCDM model. 
The number of particles of both
gas, dark matter and star particles was varied by more than an order of
magnitude. 
%In all
%runs we adopted the standard Monaghan artificial viscosity with the 
%reduction term for shear flows introduced by Balsara (1995).  
The relative angular momentum loss in the disk
for the different runs is shown in Figure 1. The plot suggests that 
 angular momentum loss mainly correlates with the number of
dark matter particles employed, as expected if two-body
heating between dark matter and star particles (most of the disk mass
is stellar) is the main responsible. 
Residual losses due to artificial viscosity and other effects [6]
may contribute to the slightly different outcomes seen in tests done 
at fixed number of dark matter particles.
The bottom line is that at least $10^5$ dark matter particles should be
used to keep artificial losses close or below the $20\%$ level (note that
we do not actually demonstrate convergence here); an equivalent, more 
useful statement is that the mass ratio between 
dark matter and disk particles should be lower than 20:1. 
The LCDM simulation in [2] satisfies this requirement at $z \sim 0.6$.

However there is another side of artificial two-body heating,
namely the increase in random kinetic energy; this will combine with
the loss of angular momentum to lower the stellar $v_{rot}/\sigma$. In Figure 2
we compare the variation of 
$v_{rot}/\sigma$ for the two test runs  which differ the
most in terms of resolution, and we also compare that with the variation
$v_{rot}/\sigma$ for the galaxy in the LCDM simulation. 
We can draw two conclusions from Figure 2; the first is that, not 
surprisigly, resolution effects on $v_{rot}/\sigma$ are even more 
dramatic than on angular momentum alone, the second is that the decrease of  
$v_{rot}/\sigma$ in the LCDM galaxy is larger than that occurring in the
test done using comparable resolution. The latter discrepancy is more evident
in the outer part of the disk; the biggest responsible for this 
is probably a large satellite, with mass roughly 1:10 of the disk mass, 
that violently plunges through the disk at $z \sim 0.1$.  We note that, taken 
at face value, the curve for the high resolution test might be telling us that 
even with $10^5$ particles there is still room for a $30-40\%$ numerical 
effect on $v_{rot}/\sigma$ (since we have not demonstrated complete
convergence yet); if that is the case it would make up for most of
the variation of $v_{rot}/\sigma$ seen in the LCDM galaxy. This is however
the worst case scenario since bar formation and buckling of the bar is
observed in the hi-res test, which certainly contributes to vertical
heating.

\section{Final remarks}

The requirement on halo mass resolution to minimize two-body effects
is not met in the early progenitors of our LCDM galaxy. 
The same is true for other
requirements on the resolution of the SPH component ([9]).
This certainly plays a role in the fact that central ``old'' 
stellar bulge tends to be too massive. While it is reasonable that 
mergers at high redshift certainly produce a kinematically hot stellar 
component, we argue that in the simulations the bulge is too massive because 
the angular momentum has been degraded by two-body heating accumulated over 
time across the merging tree ([13]).
Seeking exact convergence for the angular momentum in a CDM model cannot
be achieved by simply increasing the mass resolution because that
would just shift the numerical loss towards earlier times in 
smaller scale progenitors of a given system that were not previously 
resolved -- there will be always a step in the merger tree that will be
modeled with too few particles.
One could imagine that if enough resolution is achieved
in the most massive progenitors the problem could be substantially solved.
Alternatively, simulations with truncated power spectra are potentially
a powerful tool to perform well-posed resolution studies since a limiting 
scale for structure formation is naturally introduced.

\end{document}